\documentclass[sigconf]{acmart}

\usepackage{enumitem}
\usepackage{graphicx} 

\AtBeginDocument{%
  \providecommand\BibTeX{{%
    \normalfont B\kern-0.5em{\scshape i\kern-0.25em b}\kern-0.8em\TeX}}}

\copyrightyear{2024}
\acmYear{2024}
\setcopyright{acmlicensed}\acmConference[IDE '24]{2024 First IDE
Workshop}{April 20, 2024}{Lisbon, Portugal}
\acmBooktitle{2024 First IDE Workshop (IDE '24), April 20, 2024, Lisbon,
Portugal}
\acmDOI{10.1145/3643796.3648467}
\acmISBN{979-8-4007-0580-9/24/04}

\acmSubmissionID{27}

\begin{document}

\title{Envisioning the Next-Generation AI Coding Assistants: \\ Insights \& Proposals}

\author{Khanh Nghiem}
\orcid{0009-0009-6962-3752}
\affiliation{%
  \institution{FPT Software AI Center}
  \country{Vietnam}
}
\email{khanhnv22@fpt.com}

\author{Anh Minh Nguyen}

\orcid{0009-0005-5091-7401}
\affiliation{%
  \institution{FPT Software AI Center}
  \country{Vietnam}
}
\email{minhna4@fpt.com}

\author{Nghi D. Q. Bui}
\orcid{0000-0003-1984-4329}
\affiliation{%
  \institution{Fulbright University}
  \country{Vietnam}
}
\email{nghi.bui@fulbright.edu.vn}


\begin{abstract}
  As a research-product hybrid group in AI for Software Engineering (AI4SE), we present four key takeaways from our experience developing in-IDE AI coding assistants. AI coding assistants should set clear expectations for usage, integrate with advanced IDE capabilities and existing extensions, use extendable backend designs, and collect app data responsibly for downstream analyses. We propose open questions and challenges that academia and industry should address to realize the vision of next-generation AI coding assistants.\end{abstract}






\maketitle

\section{Introduction}
In this paper, we discuss the insights gained from our unique position conducting research and building products in the field of AI for Software Engineering (AI4SE). Our academic segment contributes to the field by publishing papers on innovative training methodologies and creating datasets tailored for code-proficient large language models. Our engineering group brings AI capabilities to developers by creating extensions and plugins for popular Integrated Development Environments (IDEs) and code editors, materializing the most recent advancements in AI4SE into developer productivity gains. Drawing from our comprehensive experience, we outline four pivotal lessons:
\begin{itemize}[leftmargin=*]
    \item AI coding assistants should clearly communicate their intended purposes and adapt feature designs accordingly.
    \item AI coding tools, IDE native capabilities, and extensions must co-evolve to provide more cohesive support to developers.
    \item Modular backend architectures allow flexible experimentation and integration of emerging AI innovations.
    \item The collection and analysis of in-app metrics and user metadata are imperative for understanding user interactions and evaluating the tools’ impact on productivity.
\end{itemize}

Through sharing our insights from combining AI4SE research and engineering, we propose where scholars and IDE developers could fruitfully invest their energy in the future.
\section{Communicate intended purposes and adapt feature design accordingly}

Despite the enormous potential of LLMs, we are still in the early stages of developing AI4SE tools that consistently produce high-quality results for specific coding tasks \cite{jimenez2023swebench} \cite{dinh2023large} \cite{fan2023large}.  
Academic researchers and industry practitioners lack well-defined frameworks for positioning and evaluating emerging AI coding assistants in the traditional programming paradigms\cite{sarkar2022like} \cite{allamanis2024unsupervised}, while users lack clear expectations of productivity gains when adopting these novel tools. This problem is compounded in enterprise environments where the vetting process for new vendors and tools is time-consuming and labor-intensive. 

We faced a similar challenge when building CodeVista, an AI coding assistant \cite{Ross_2023} that provides a conversational interface and predefined workflows to assist with various coding tasks. It uses several instruction-tuned AI models, including OpenAI GPT-3.5 and GPT-4 \cite{openai2023gpt4}, Meta LLaMA 2 \cite{rozière2024code}, Google Code Chat Bison, and Google Gemini \cite{geminiteam2023gemini}. Pilot users were perplexed about how to effectively interact with CodeVista. Users who were unfamiliar with prompt engineering frequently wrote incomplete utterances when performing complex tasks like generating or refactoring an entire program. To address this behavior, we created a library of sample prompts for different coding tasks as a reference and established predefined workflows to guide user expectations. For example, the “code explain” divides a file into structural blocks before reasoning through each component, summarizing the overall logic of the file, and highlighting potential issues. Google Search integration allows users to look up similar implementations or sources about coding concepts or libraries mentioned in the explanation. Our team defined this workflow by focusing on a single use case and brainstorming the design with pilot users.

\begin{figure}[h]
  \centering
  \includegraphics[width=\linewidth]{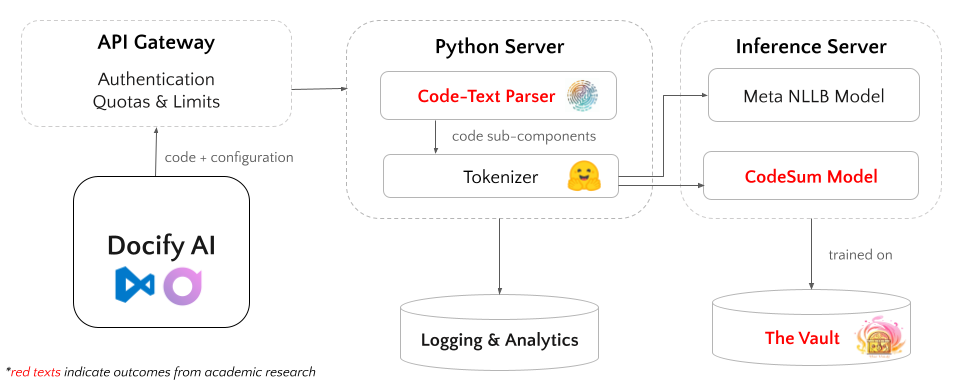}
  \caption{Docify Architecture}
  \Description{Architecture of Docify}
  \label{fig:docify}
\end{figure}

For "code documentation", we integrated Docify AI (Figure \ref{fig:docify}), our VSCode extension for writing docstrings and code comments from source code that we built on top of CodeSum, our proprietary code-to-text AI model. To create CodeSum, we applied research outcomes from our lab, including the knowledge distillation training architecture \cite{to2023better} and TheVault \cite{manh2023vault} – currently the largest code-text pair dataset. Docify AI enables two specific use cases for users: (1) to explore legacy code bases that lack proper documentation, and (2) to increase documentation coverage and reduce manual effort in coding projects. These clear-cut scenarios help set user expectations and guide the engineering team’s product design strategies. 

From our experience and literature review in the Human-Computer Interaction field, we observe that users face confusion when using general-purpose AI4SE tools \cite{zhou2023concerns}\cite{kazemitabaar2023novices}. We call on both academia and industry to conduct further user studies and devise guidelines for application designs that will bring more utility and positive experience for users \cite{wang2023investigating}.

\section{Co-evolve AI coding assistants with IDE native capabilities and extensions}
Effective AI coding assistants should fit smoothly into developers’ workflows. Achieving this means aligning the functionalities of AI coding assistants with the build-in features of IDEs for similar tasks. For instance, we designed CodeVista to act as a pair programmer that suggests edits to existing code. To enhance readability and user control over AI-generated code, we use the VSCode refactor preview panel as a visual prompt for users to review CodeVista’s suggestions carefully before applying any changes. However, there are times when the standard features of an IDE don’t directly correspond to what AI4SE tools aim to accomplish. In such scenarios, our engineering team has to get creative. Take Docify AI as an example: it introduces ICommentBlock components in VSCode, serving as “annotations” that allow for previewing and watermarking AI-generated docstrings and comments directly in the editor.

IDE developers are encouraged to make native features more accessible to AI4SE tools. A case in point is GitHub Copilot Chat, which enables users to start inline discussions from selected code in the editor. Through reverse engineering, we discovered the need for 10 specific VSCode Proposed APIs to create this interactive experience. While these APIs are available to extension developers, restrictions exist on distributing extensions using them on the Marketplace, with the exception of GitHub Copilot Chat. As a workaround, CodeVista employs the stable createCommentController API from VSCode to craft a functional, though makeshift, user interface. In addition to tapping into IDE’s native capabilities, AI4SE tools should also complement existing static analysis extensions. For example, CodeVista works alongside the SonarLint extension in VSCode to facilitate code refactoring. It uses SonarLint’s analysis of code convention breaches within the same workspace to inform the AI coding assistant to provide code improvement suggestions.

\section{Enable experimentation and innovations through engineering}

\begin{figure}[h]
  \centering
  \includegraphics[width=\linewidth]{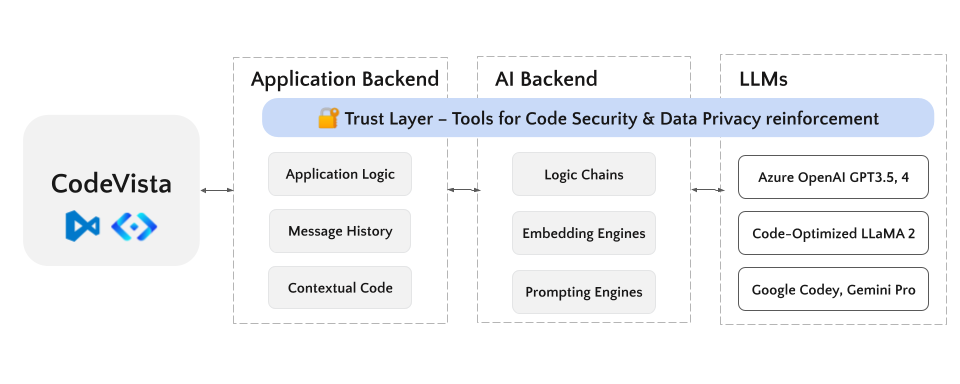}
  \caption{CodeVista Architecture}
  \Description{Architecture of CodeVista}
  \label{fig:codevista}
\end{figure}

AI4SE is evolving quickly with new AI models, frameworks, and application libraries released frequently. To remain cutting-edge, we suggest IDE developers employ extendable, loosely coupled architectures to adapt to fast-changing technologies. We present CodeVista’s architecture as a reference
 (Figure \ref{fig:codevista}). 

Separation of concerns enables team members to independently experiment with new technologies and ship new features. For example, there are multiple stateless deployments of the AI backend servers, one for each AI model (GPT 3.5, GPT 4, LLaMA 2, Gemini Pro, Code Chat, etc.) that scale individually. State management is proxied through the Application backend layer, including conversation history, few-shot adaptations, and user configurations. Therefore, AI backend engineers can quickly experiment with new updates from prompting frameworks such as LangChain or inference frameworks such as HuggingFace TGI and Ray, without any cascading effects on the conversation management component. This architecture also enables end users to switch seamlessly from one AI model to another AI model in the middle of the same conversation.

The most important innovation project that our research and engineering teams are collaborating on is enabling CodeVista to perform repository-level coding tasks, including code refactoring and systematic debugging. These tasks require:
\begin{itemize}[leftmargin=*]
  \item  Static analysis tool (call graph, data flow graph, change management, retrieval) and multi-turn coordination, which fall within the scope of the Application backend engineering team,
  \item Multi-agent framework \cite{qian2023communicative} for task planning, execution, and verification, which falls within the scope of the AI backend engineering team and academic research group.

\end{itemize}

The loosely coupled architecture allows our team to remain nimble in experimenting with and implementing the building blocks of this system. As our team and others build proof-of-concept implementations of these systems, we should share the design guidelines and critical analyses to quickly materialize the vision of AI-automated coding systems.

\section{Collect app data to understand user behavior and impact analysis}
Since we are still in the early stages of implementing AI coding assistants, it is crucial to collect and study data in a transparent, responsible manner that respects privacy. Currently, CodeVista encrypts conversation history to ensure no one except the conversation owner can uncover the content. For downstream analyses, we only collect metadata such as the length in the token count of user inputs including user messages and context code snippets, along with user ratings (helpful or unhelpful) of CodeVista-generated responses and user actions (e.g., inserting CodeVista-generated code into the active editor). To enrich available data, we generate the user intent \cite{shah2023using} for each request based on a taxonomy of 20 coding operations using GPT-3.5. These data help us track trends in user usage, estimate productivity gains, and improve few-shot adaptations \cite{su2022selective}. 

Although we see emerging methods for quantifying and reasoning AI impact on developer productivity, these often lack generality and scalability. We call on both academia and industry to conduct further research on frameworks and tools that can shed light on how our AI4SE community can better serve people’s coding productivity and experience.
\bibliographystyle{ACM-Reference-Format}
\bibliography{main}


\begin{thebibliography}{17}


\ifx \showCODEN    \undefined \def \showCODEN     #1{\unskip}     \fi
\ifx \showDOI      \undefined \def \showDOI       #1{#1}\fi
\ifx \showISBNx    \undefined \def \showISBNx     #1{\unskip}     \fi
\ifx \showISBNxiii \undefined \def \showISBNxiii  #1{\unskip}     \fi
\ifx \showISSN     \undefined \def \showISSN      #1{\unskip}     \fi
\ifx \showLCCN     \undefined \def \showLCCN      #1{\unskip}     \fi
\ifx \shownote     \undefined \def \shownote      #1{#1}          \fi
\ifx \showarticletitle \undefined \def \showarticletitle #1{#1}   \fi
\ifx \showURL      \undefined \def \showURL       {\relax}        \fi
\providecommand\bibfield[2]{#2}
\providecommand\bibinfo[2]{#2}
\providecommand\natexlab[1]{#1}
\providecommand\showeprint[2][]{arXiv:#2}

\bibitem[AI(2024)]%
        {rozière2024code}
\bibfield{author}{\bibinfo{person}{Meta AI}.} \bibinfo{year}{2024}\natexlab{}.
\newblock \bibinfo{title}{Code Llama: Open Foundation Models for Code}.
\newblock
\newblock
\showeprint[arxiv]{2308.12950}~[cs.CL]


\bibitem[Allamanis et~al\mbox{.}(2024)]%
        {allamanis2024unsupervised}
\bibfield{author}{\bibinfo{person}{Miltiadis Allamanis}, \bibinfo{person}{Sheena Panthaplackel}, {and} \bibinfo{person}{Pengcheng Yin}.} \bibinfo{year}{2024}\natexlab{}.
\newblock \bibinfo{title}{Unsupervised Evaluation of Code LLMs with Round-Trip Correctness}.
\newblock
\newblock
\showeprint[arxiv]{2402.08699}~[cs.SE]


\bibitem[Dinh et~al\mbox{.}(2023)]%
        {dinh2023large}
\bibfield{author}{\bibinfo{person}{Tuan Dinh}, \bibinfo{person}{Jinman Zhao}, \bibinfo{person}{Samson Tan}, \bibinfo{person}{Renato Negrinho}, \bibinfo{person}{Leonard Lausen}, \bibinfo{person}{Sheng Zha}, {and} \bibinfo{person}{George Karypis}.} \bibinfo{year}{2023}\natexlab{}.
\newblock \bibinfo{title}{Large Language Models of Code Fail at Completing Code with Potential Bugs}.
\newblock
\newblock
\showeprint[arxiv]{2306.03438}~[cs.LG]


\bibitem[Fan et~al\mbox{.}(2023)]%
        {fan2023large}
\bibfield{author}{\bibinfo{person}{Angela Fan}, \bibinfo{person}{Beliz Gokkaya}, \bibinfo{person}{Mark Harman}, \bibinfo{person}{Mitya Lyubarskiy}, \bibinfo{person}{Shubho Sengupta}, \bibinfo{person}{Shin Yoo}, {and} \bibinfo{person}{Jie~M. Zhang}.} \bibinfo{year}{2023}\natexlab{}.
\newblock \bibinfo{title}{Large Language Models for Software Engineering: Survey and Open Problems}.
\newblock
\newblock
\showeprint[arxiv]{2310.03533}~[cs.SE]


\bibitem[Jimenez et~al\mbox{.}(2023)]%
        {jimenez2023swebench}
\bibfield{author}{\bibinfo{person}{Carlos~E. Jimenez}, \bibinfo{person}{John Yang}, \bibinfo{person}{Alexander Wettig}, \bibinfo{person}{Shunyu Yao}, \bibinfo{person}{Kexin Pei}, \bibinfo{person}{Ofir Press}, {and} \bibinfo{person}{Karthik Narasimhan}.} \bibinfo{year}{2023}\natexlab{}.
\newblock \bibinfo{title}{SWE-bench: Can Language Models Resolve Real-World GitHub Issues?}
\newblock
\newblock
\showeprint[arxiv]{2310.06770}~[cs.CL]


\bibitem[Kazemitabaar et~al\mbox{.}(2023)]%
        {kazemitabaar2023novices}
\bibfield{author}{\bibinfo{person}{Majeed Kazemitabaar}, \bibinfo{person}{Xinying Hou}, \bibinfo{person}{Austin Henley}, \bibinfo{person}{Barbara~J. Ericson}, \bibinfo{person}{David Weintrop}, {and} \bibinfo{person}{Tovi Grossman}.} \bibinfo{year}{2023}\natexlab{}.
\newblock \bibinfo{title}{How Novices Use LLM-Based Code Generators to Solve CS1 Coding Tasks in a Self-Paced Learning Environment}.
\newblock
\newblock
\showeprint[arxiv]{2309.14049}~[cs.HC]


\bibitem[Manh et~al\mbox{.}(2023)]%
        {manh2023vault}
\bibfield{author}{\bibinfo{person}{Dung~Nguyen Manh}, \bibinfo{person}{Nam~Le Hai}, \bibinfo{person}{Anh T.~V. Dau}, \bibinfo{person}{Anh~Minh Nguyen}, \bibinfo{person}{Khanh Nghiem}, \bibinfo{person}{Jin Guo}, {and} \bibinfo{person}{Nghi D.~Q. Bui}.} \bibinfo{year}{2023}\natexlab{}.
\newblock \bibinfo{title}{The Vault: A Comprehensive Multilingual Dataset for Advancing Code Understanding and Generation}.
\newblock
\newblock
\showeprint[arxiv]{2305.06156}~[cs.CL]


\bibitem[OpenAI(2023)]%
        {openai2023gpt4}
\bibfield{author}{\bibinfo{person}{OpenAI}.} \bibinfo{year}{2023}\natexlab{}.
\newblock \bibinfo{title}{GPT-4 Technical Report}.
\newblock
\newblock
\showeprint[arxiv]{2303.08774}~[cs.CL]


\bibitem[Qian et~al\mbox{.}(2023)]%
        {qian2023communicative}
\bibfield{author}{\bibinfo{person}{Chen Qian}, \bibinfo{person}{Xin Cong}, \bibinfo{person}{Wei Liu}, \bibinfo{person}{Cheng Yang}, \bibinfo{person}{Weize Chen}, \bibinfo{person}{Yusheng Su}, \bibinfo{person}{Yufan Dang}, \bibinfo{person}{Jiahao Li}, \bibinfo{person}{Juyuan Xu}, \bibinfo{person}{Dahai Li}, \bibinfo{person}{Zhiyuan Liu}, {and} \bibinfo{person}{Maosong Sun}.} \bibinfo{year}{2023}\natexlab{}.
\newblock \bibinfo{title}{Communicative Agents for Software Development}.
\newblock
\newblock
\showeprint[arxiv]{2307.07924}~[cs.SE]


\bibitem[Ross et~al\mbox{.}(2023)]%
        {Ross_2023}
\bibfield{author}{\bibinfo{person}{Steven~I. Ross}, \bibinfo{person}{Fernando Martinez}, \bibinfo{person}{Stephanie Houde}, \bibinfo{person}{Michael Muller}, {and} \bibinfo{person}{Justin~D. Weisz}.} \bibinfo{year}{2023}\natexlab{}.
\newblock \showarticletitle{The Programmer’s Assistant: Conversational Interaction with a Large Language Model for Software Development}. In \bibinfo{booktitle}{\emph{Proceedings of the 28th International Conference on Intelligent User Interfaces}} \emph{(\bibinfo{series}{IUI ’23})}. \bibinfo{publisher}{ACM}.
\newblock
\urldef\tempurl%
\url{https://doi.org/10.1145/3581641.3584037}
\showDOI{\tempurl}


\bibitem[Sarkar et~al\mbox{.}(2022)]%
        {sarkar2022like}
\bibfield{author}{\bibinfo{person}{Advait Sarkar}, \bibinfo{person}{Andrew~D. Gordon}, \bibinfo{person}{Carina Negreanu}, \bibinfo{person}{Christian Poelitz}, \bibinfo{person}{Sruti~Srinivasa Ragavan}, {and} \bibinfo{person}{Ben Zorn}.} \bibinfo{year}{2022}\natexlab{}.
\newblock \bibinfo{title}{What is it like to program with artificial intelligence?}
\newblock
\newblock
\showeprint[arxiv]{2208.06213}~[cs.HC]


\bibitem[Shah et~al\mbox{.}(2023)]%
        {shah2023using}
\bibfield{author}{\bibinfo{person}{Chirag Shah}, \bibinfo{person}{Ryen~W. White}, \bibinfo{person}{Reid Andersen}, \bibinfo{person}{Georg Buscher}, \bibinfo{person}{Scott Counts}, \bibinfo{person}{Sarkar Snigdha~Sarathi Das}, \bibinfo{person}{Ali Montazer}, \bibinfo{person}{Sathish Manivannan}, \bibinfo{person}{Jennifer Neville}, \bibinfo{person}{Xiaochuan Ni}, \bibinfo{person}{Nagu Rangan}, \bibinfo{person}{Tara Safavi}, \bibinfo{person}{Siddharth Suri}, \bibinfo{person}{Mengting Wan}, \bibinfo{person}{Leijie Wang}, {and} \bibinfo{person}{Longqi Yang}.} \bibinfo{year}{2023}\natexlab{}.
\newblock \bibinfo{title}{Using Large Language Models to Generate, Validate, and Apply User Intent Taxonomies}.
\newblock
\newblock
\showeprint[arxiv]{2309.13063}~[cs.IR]


\bibitem[Su et~al\mbox{.}(2022)]%
        {su2022selective}
\bibfield{author}{\bibinfo{person}{Hongjin Su}, \bibinfo{person}{Jungo Kasai}, \bibinfo{person}{Chen~Henry Wu}, \bibinfo{person}{Weijia Shi}, \bibinfo{person}{Tianlu Wang}, \bibinfo{person}{Jiayi Xin}, \bibinfo{person}{Rui Zhang}, \bibinfo{person}{Mari Ostendorf}, \bibinfo{person}{Luke Zettlemoyer}, \bibinfo{person}{Noah~A. Smith}, {and} \bibinfo{person}{Tao Yu}.} \bibinfo{year}{2022}\natexlab{}.
\newblock \bibinfo{title}{Selective Annotation Makes Language Models Better Few-Shot Learners}.
\newblock
\newblock
\showeprint[arxiv]{2209.01975}~[cs.CL]


\bibitem[Team(2023)]%
        {geminiteam2023gemini}
\bibfield{author}{\bibinfo{person}{Gemini Team}.} \bibinfo{year}{2023}\natexlab{}.
\newblock \bibinfo{title}{Gemini: A Family of Highly Capable Multimodal Models}.
\newblock
\newblock
\showeprint[arxiv]{2312.11805}~[cs.CL]


\bibitem[To et~al\mbox{.}(2023)]%
        {to2023better}
\bibfield{author}{\bibinfo{person}{Hung~Quoc To}, \bibinfo{person}{Nghi D.~Q. Bui}, \bibinfo{person}{Jin Guo}, {and} \bibinfo{person}{Tien~N. Nguyen}.} \bibinfo{year}{2023}\natexlab{}.
\newblock \bibinfo{title}{Better Language Models of Code through Self-Improvement}.
\newblock
\newblock
\showeprint[arxiv]{2304.01228}~[cs.CL]


\bibitem[Wang et~al\mbox{.}(2023)]%
        {wang2023investigating}
\bibfield{author}{\bibinfo{person}{Ruotong Wang}, \bibinfo{person}{Ruijia Cheng}, \bibinfo{person}{Denae Ford}, {and} \bibinfo{person}{Thomas Zimmermann}.} \bibinfo{year}{2023}\natexlab{}.
\newblock \bibinfo{title}{Investigating and Designing for Trust in AI-powered Code Generation Tools}.
\newblock
\newblock
\showeprint[arxiv]{2305.11248}~[cs.HC]


\bibitem[Zhou et~al\mbox{.}(2023)]%
        {zhou2023concerns}
\bibfield{author}{\bibinfo{person}{Xiyu Zhou}, \bibinfo{person}{Peng Liang}, \bibinfo{person}{Beiqi Zhang}, \bibinfo{person}{Zengyang Li}, \bibinfo{person}{Aakash Ahmad}, \bibinfo{person}{Mojtaba Shahin}, {and} \bibinfo{person}{Muhammad Waseem}.} \bibinfo{year}{2023}\natexlab{}.
\newblock \bibinfo{title}{On the Concerns of Developers When Using GitHub Copilot}.
\newblock
\newblock
\showeprint[arxiv]{2311.01020}~[cs.SE]


\end{thebibliography}

\end{document}